\pdfoutput=1

\documentclass[acmsmall,nonacm,dvipsnames]{acmart} 
\renewcommand\footnotetextcopyrightpermission[1]{} 
\usepackage[utf8]{inputenc}

\usepackage{booktabs} 
\usepackage{amsmath,amssymb}
\usepackage{soul}
\usepackage{subcaption}
\usepackage{xcolor}
\usepackage{tikz}
\usetikzlibrary{arrows,positioning,spy,shapes.geometric,calc}
\usepackage{ifthen}
\usepackage{graphicx}
\usepackage[export]{adjustbox}
\usepackage{arydshln}
\usepackage{multirow}

\tikzset{
    agent/.style={
            rectangle, rounded corners,
            draw=black, very thick,
            text width=6em, text centered,
            minimum height=2em
    },
    fusion-agent/.style={
            agent,
            draw = ForestGreen
    },
    prediction-agent/.style={
            agent,
            draw = RoyalBlue
    },
    check-agent/.style={
            agent,
            draw = BrickRed
    },
    source-agent/.style={
            agent,
            circle,
            text width=3.2em
    },
    database/.style={
      cylinder,
      cylinder uses custom fill,
      cylinder body fill=white,
      cylinder end fill=white,
      shape border rotate=90,
      aspect=0.15,
      draw,
      minimum width = 5em
    },
    process/.style={
      draw,
      black,
      rounded corners
    },
    >=stealth',
    stream/.style={
            ->,
            very thick,
            shorten >=1pt
    }
}

\usepackage{mdframed}

\newcommand{\capsule}{{\sc CaPSuLe}}

\title{The Assurance Monitor Pattern}

\author[Duracz]{Adam Duracz}
\affiliation{~Rice University}
\email{adam.duracz@rice.edu}

\author[Chandy]{K. Mani Chandy}
\affiliation{~Caltech}
\email{kmchandy@gmail.com}

\author[Abdelrahman]{Mohamed Abdelrahman}
\affiliation{~Rice University}
\email{mhafez@rice.edu}

\author[España]{Juan Jose Gonzalez España}
\affiliation{~Rice University}
\email{jose.gonzalez@rice.edu}

\author[Sai]{Ryuichi Sai}
\affiliation{~Rice University}
\email{ryuichi@rice.edu}

\author[Yang]{Yao-Hsiang Yang}
\affiliation{~Rice University}
\email{yao-hsiang.yang@rice.edu}

\author[Cartwright]{Robert Cartwright}
\affiliation{~Rice University}
\email{cork@rice.edu}

\author[Palem]{Krishna V. Palem}
\affiliation{~Rice University}
\email{palem@rice.edu}

\date{February 2018}

\begin{abstract}
Some applications require an assurance that certain criteria are
violated with only low probability. An alert is generated when the
current course of action is likely to violate assurance criteria and
the alert results in corrective action. Assurance monitors fuse
information from multiple data streams generated by sensors and other sources to estimate the probability distribution of
system trajectories. This distribution is used to determine whether an
assurance constraint is likely to be violated. The acquisition of data
requires resources such as energy, computational power and communication 
bandwidth which are scarce in many applications. At each point in time the
system must decide whether to expend these resources to get more data
to improve the confidence in the probability distribution of
trajectories. 
This paper presents a software-design pattern for assurance monitoring and 
gives examples of its uses, including an application of the pattern to the 
problem of autonomous navigation of a drone which is required to avoid 
no-fly zones while using limited resources.
\end{abstract}

\begin{document}
    

\maketitle

\thispagestyle{empty}

\section*{Introduction}

Software design patterns entered the vocabulary of the broad software engineering community with the publication of the classic book by Gamma, Vlissides, Johnson, and  Helm~\cite{gamma1995patterns}. The concept provided systems designers with a convenient way of describing solutions to common problems, along with an analysis of the consequences of adopting the prescribed solutions. More than two decades after its publication, the concept remains influential to the way software engineers think about problem solving. Today, design patterns have been adopted across a variety of domains beyond object-oriented design including security~\cite{fernandez2001pattern}, generic programming~\cite{lammel2003scrap}, graphics~\cite{heer2006software} and distributed data processing~\cite{kiran2015lambda}, demonstrating the general appeal of the concept.

Though the term was originally borrowed from architecture~\cite{alexander1977pattern}, designing systems by composing well-understood elements is inherent to the practice of engineering. For example, control systems that use feedback can robustly maintain their behaviour under state changes, and controller patterns such as proportional-integral-derivative (PID) are used to construct systems with well-understood behavior. 

\begin{figure}[htpb]
\begin{minipage}[b]{0.47\textwidth}
    \centering
    \small
    \begin{tikzpicture}
        \node[anchor=south west, inner sep=0] (terrain) at (0,0) {
            \includegraphics[width=\linewidth]{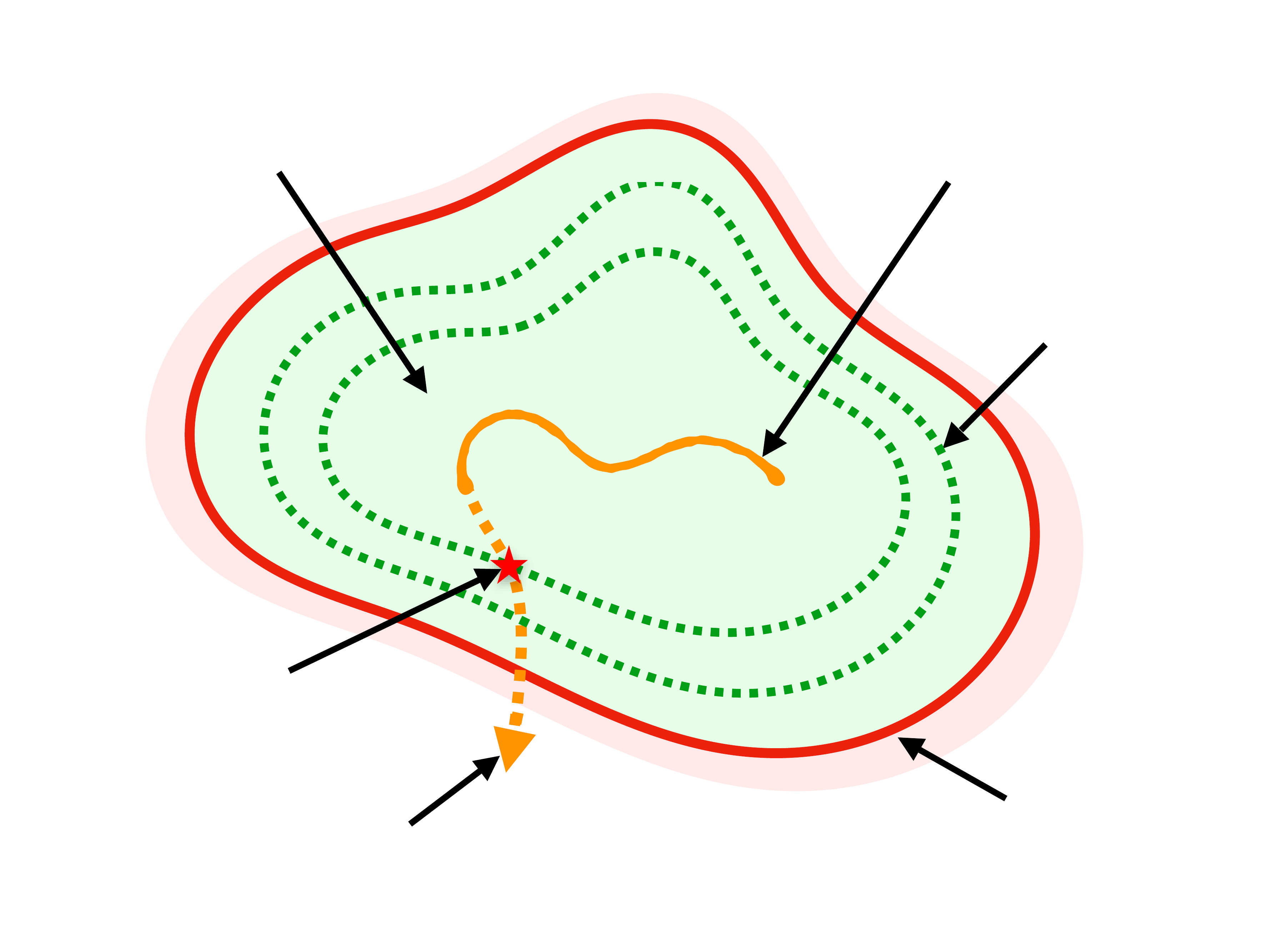}
        };
        \node[align=center] at (0.6,4.4)
            {Safe region of\\$n$-dimensional\\state space};
        \node[align=center] at (4.9,4.4)
            {Actual\\trajectory};
        \node[align=center] at (5.6,3.55)
            {Safety\\margin};
        \node[align=center] at (0.7,1.4)
            {Assurance\\violation};
        \node[align=center] at (1.35,0.5)
            {Predicted\\trajectory};
        \node[align=center] at (5.95,0.8)
            {Safe region\\boundary};
    \end{tikzpicture}
    \caption{Assurance Monitoring of a State Trajectory}
    \label{fig:state-space}
\end{minipage}
\hfill
\begin{minipage}[b]{0.47\textwidth}
    \centering
    \begin{tikzpicture}[scale=0.85]
        \foreach \x in {0,...,7}
            \foreach \y in {0,...,5} 
                \coordinate (\x\y) at (1.1*\x,0.6*\y);
        \foreach \x in {1,...,6}  
            \draw[dashed,RoyalBlue,thick] 
                ($ (\x0) - ( 0.0, -0.25) $) -- ($ (\x4) + ( 0.0, 0.45) $);
        \foreach \y in {1,...,4}  
            \draw[dashed,RoyalBlue,thick] 
                ($ (0\y) - (-0.5, 0.0) $) -- ($ (6\y) + ( 0.5, 0.0) $);
        \foreach \x  in {2,...,6}
            \foreach \y in {1,...,3}
            {
                \draw[ ->,red,thick
                     , shorten >=3, shorten <=3
                     ] 
                    ($(7*1.1,5*0.6)-(\x\y)$) -- 
                    ($(7*1.1,5*0.6)-(\x\y)+(1.1,-0.6)$);
            }
        \foreach \x  in {2,...,6}
            \foreach \y in {1,...,2}
            {
                \draw[ ->,red,thick
                     , shorten >=3, shorten <=3
                     ] 
                    ($(7*1.1,5*0.6)-(\x\y)$) -- 
                    ($(7*1.1,5*0.6)-(\x\y)+(1.1,-0.6*2)$);
            }
        \foreach \x  in {2,...,6}
            \foreach \y in {1}
            {
                \draw[ ->,red,thick
                     , shorten >=3, shorten <=3
                     ] 
                    ($(7*1.1,5*0.6)-(\x\y)$) -- 
                    ($(7*1.1,5*0.6)-(\x\y)+(1.1,-0.6*3)$);
            }
        \foreach \x  in {1,...,6}
            \foreach \y in {1,...,4}  
                \draw[fill=white,thick] (\x\y) circle (0.075);
        \node[align=center, above = 0em of 15] {Mission\\starts};
        \node[align=center, above = 0em of 35] {Current\\time};
        \node[align=center, above = 0em of 65] {Mission\\ends};
        \coordinate[below=0em of 10] (timeStart);
        \coordinate[below=0em of 60] (timeEnd);
        \draw[->] (timeStart) -- (timeEnd) node[pos=0.5,below] {Time};
    \end{tikzpicture}
    \caption{
        Continuous prediction of probability of assurance violation.
        Red arrows represent possible transitions between states.
    }
    \label{fig:prediction}
\end{minipage}
\end{figure}

Patterns play a particularly important role in distributed systems, whose design may be hard to change after they are deployed. Patterns such as communication over queues, message brokering~\cite{hohpe2004enterprise}, and publish/subscribe~\cite{birman1987exploiting} provide systems with desirable qualities such as fault tolerance, scalability, loose coupling and guaranteed-once delivery~\cite{erl2008soa}.
Enterprise integration has long benefited from the availability of software platforms that support such patterns. Novel patterns are being discovered in several emerging domains that involve event-based systems. In cloud computing, immutable data storage~\cite{apache-kafka} addresses the combined challenge of scale and latency. In IoT, shadowing and triggering~\cite{reinfurt2017internet} address intermittent availability of networked, energy-constrained devices, and metering~\cite{qanbari2016iot} enables associated business models. In cyber-physical systems, the way in which these systems are deployed and interact with one another gives rise to patterns to deal with time and synchronization~\cite{graja2016time} and the need to adapt to changes in resources and the operating environment~\cite{musil2017patterns}.

Patterns can help to get more widespread adoption of 
event-based systems (EBS) ideas and software. A person developing an EBS
application may not be able to use a program written by someone else
exactly as the program is written. This could be
because the programming language in the earlier application is
different from the new one, or the libraries that
the earlier application used may be different from those that the
developer wants to use (e.g. MATLAB versus Python
SciPy Toolkits). Nevertheless, many EBS applications have similar constructs,
and often these constructs can be represented as reusable
patterns. This paper proposes a pattern for applications that monitor
and, if necessary, take steps to correct, the behavior of systems.

The structure of the pattern can be represented by a graph in which
the vertices are agents and the edges represent information flow. The
behavior of an agent is that it is asleep and not carrying out any
computation until it is woken up by a clock or by the arrival of
information from another agent. The agent can ``pull'' information;
for example it can call a GPS sensor when it wants a GPS
reading. Also, information can be ``pushed'' to agents; for example,
an agent may receive a stream of accelerations. This structure has
been described since the inception of EBS, and books on EBS
\cite{luckham2002power,chandy2009event,etzion2011event} describe this pattern in one form or
another.

The contribution of this paper is to use this structure to form a
pattern that is used in a particular kind of EBS application:
assurance monitoring. The pattern uses agents for Bayesian fusion of
information from multiple sources, predicting future system behavior,
and checking whether future behavior is likely to violate
constraints. The way in which agents are connected can vary from
application to application though all assurance monitors have the same
essential structure to deal with the question: Is future
behavior safe? The pattern is used for a specific application of drone
navigation, which is discussed in detail; then the use of the same
pattern in other applications is described briefly. An implementation
of the pattern, in Python, is freely available~\cite{assurance-monitor-poc-source}.

The assurance monitor pattern arises often in event-based systems and, in particular, addresses the need to predict undesirable behavior in autonomous systems, whose behavior is difficult to analyze. 
Such systems are currently being introduced across all of the aforementioned domains, spurred by recent advances in machine learning.
The rest of this paper describes the assurance monitor pattern (Section~\ref{sec:assurance-monitor-pattern}), describes an instance of the pattern implemented in Python (Section~\ref{sec:drone-navigation}), discusses experimental results based on this implementation (Section~\ref{sec:experiment}), and concludes with a discussion of its use in a second application (Section~\ref{sec:another-example}).

\section{Assurance Monitoring}

Figure~\ref{fig:state-space} illustrates the general problem solved by an assurance monitor, in terms of a system state space that consists of a safe region, where the system is intended to operate, and an unsafe region, where there is some risk of harm. The assurance monitor predicts the future trajectory of the system, and estimates when the probability of entering the unsafe region exceeds a threshold, which corresponds to a safety margin around the unsafe region.
Assurance monitors can observe multiple thresholds, which may correspond to different behavior such as a warning, a request for more information, or a corrective action.

The state space of a system can include both discrete and continuous dimensions. Discrete states arise from computational aspects of the system's behavior, and include the number/type of processor cores or amount of memory available to distribute the system's computations over, and the maximal clock rate that can be utilized without overheating the system.
Continuous dimensions arise from the physical aspects of the system, and include space, time, temperature, voltage and current.

Generally, to make computational analysis of such hybrid (discrete/continuous) systems tractable, continuous dimensions are discretized by dividing the space into a finite number of regions. One thus obtains a discrete representation of the system that can be reasoned about using traditional techniques.
For example, by assuming that each state is a time-dependent random variable we obtain Markov model that, given a probability distribution over the system states at the current time, and probabilities for transitioning between each of the discrete states, can be used to predict future system state probability distributions. This process is illustrated in Figure~\ref{fig:prediction}, where each time instant is represented by a column of discrete states, with possible transitions from states at the current time contributing to the probability of states at the next time.

The problem of assurance monitoring of autonomous systems can be
complex. Here we point out a few of the factors that make the problem
difficult in the context of autonomous drone navigation. 
The no-fly zone can be a complex structure; if no-fly zones
consist of densely populated areas in an urban region then the zones
can form complex geometrical shapes. The optimal use of scarce
resources is difficult in these situations because resources when
flying over simpler shapes may have to be saved for later use, when
flying over complex structures.  One approach is to use a
finite-horizon dynamic program and particle filters. This paper only
considers simple no-fly geometries, does not use dynamic programs, and
limits decision-making algorithms to simple heuristics.

A problem with visual navigation (using photographs) is that the drone may
be estimated, with high probability, to be in multiple locations that
are far apart. Consider the case of a drone flying east which is
estimated, with high probability, to be in location $A$ and with equal
probability to be in a distant location $B$. If a no-fly zone abuts $A$
to the north and a no-fly zone abuts $B$ to the south, then optimal
control can be complex. In this paper, we assume that the drone is in
the (single) location with the maximum likelihood probability.

Much more work needs to be done on the code developed so far for use
in visual autonomous navigation. We believe, however, that our continued
development of the application will continue to use assurance
monitoring patterns. Moreover, the pattern can be used
with very different applications. As earlier papers and books on
event processing have suggested implicitly, the use of EBS patterns
can help the wider community develop event-processing applications.

\section{An Assurance Monitor Pattern}
\label{sec:assurance-monitor-pattern}

Some event-based applications require an assurance that certain criteria are satisfied at all times or are violated with only low
probability. An alert is generated when the current course of action
is likely to violate assurance criteria. The alert results in
corrective action. This pattern concerns the detection of events that
signal the probable violation of assurance criteria. The pattern can be summarized as follows:\\
{\vspace{0.5em}\begin{mdframed}[
     innertopmargin=0.3em,
     rightline=false,
     leftline=false
]\itshape
\vspace{0.3em}
Given limited knowledge about the behavior of an event-based application, avoid violating an assurance criterion by predicting future events, using data from multiple sources to improve the accuracy of the prediction.
\vspace{0.3em}
\end{mdframed}
\vspace{0.5em}}

\subsection{Intent and Motivation} 
Assurance monitors fuse information from multiple data streams
generated by sensors and other sources to estimate the
probability distribution of system trajectories. This distribution is
used to determine whether an assurance constraint is likely to be
violated. The acquisition of data may require resources such as energy, 
computational power or communication bandwidth which are scarce in many
applications, or incur charges which should be kept to a minimum. 
At each point in time the system must decide whether to
expend these resources to get more data to improve the confidence in
the probability distribution of trajectories.

Informally speaking, the system should not expend its resource budget
to get more data when it is confident that the current trajectory will
not violate assurance criteria for a specified time horizon.  If,
however, the system is not confident that the assurance criteria can
be maintained, then if adequate resources are available the system
should spend part of its resource budget to get additional data to get
a better estimate of the state probability distribution.  If the
system is not confident that the assurance criteria can be maintained
and its resources are depleted then it should generate an alert
signaling the need for a course correction. The assurance monitor
balances the need for ensuring that constraints are not violated while
ensuring that resources are not depleted.

\subsection{Structure}

\begin{figure}
    \tikzset{
        stream-tuple/.style={
                ->,
                very thick,
                shorten >=1pt,
                dashed
        },
        stream-feedback/.style={
                ->,
                very thick,
                shorten >=1pt,
                dotted
        },
        stream-feedback-noarrow/.style={
                very thick,
                shorten >=1pt,
                dotted
        }
    }
    \newsavebox{\legendbox}
    \sbox{\legendbox}{%
        \begin{tikzpicture}
            \node (TupleSource) {}; 
            \node [ right = 3em of TupleSource
                  , align = left
                  ] 
                (TupleSink) 
                {Stream of: (Trajectory Distribution, Control)};
            \node [below = 1em of TupleSource
                  ]
                  (FeedbackSource) {};
            \node [right = 3em of FeedbackSource
                  ]
                  (FeedbackSink) 
                  {Stream of: (Trajectory Distribution, Feedback)};
            \draw[stream-tuple]    (TupleSource) -- (TupleSink);
            \draw[stream-feedback] (FeedbackSource) -- (FeedbackSink);
        \end{tikzpicture}
    }
    \centering
    \begin{tikzpicture}[node distance=1.5cm]
    \node [prediction-agent] 
        (Predict) 
        {Predict};
    \node [ fusion-agent
          , below of = Predict
          ] 
        (Fusion) 
        {Fuse \\ Measurements};
    \node [ source-agent
          , right = 2em of Fusion
          , xshift =  0.2em
          , yshift = -0.2em
          ]
        (Source2) 
        {Sources};
    \node [ source-agent
          , right = 2em of Fusion
          , xshift =  0.4em
          , yshift = -0.4em
          ]
        (Source3) 
        {Sources};
    \node [ source-agent
          , right = 2em of Fusion
          , fill=white
          ]
        (Source) 
        {Sources};
    \node [ check-agent
          , below of = Fusion
          ]
        (Check) 
        {Check Violation};
    \node [ agent
          , left = 2em of Fusion
          , align=center
          ]
        (Act) 
        { Control
        };
    \coordinate [below of = Act] (join) {};
    \draw[stream-tuple] 
        (Predict) --
        (Fusion);
    \draw[stream] 
        (Source) --
        (Fusion);
    \draw[stream-tuple] 
        (Fusion) --
        (Check);
    \draw[stream-feedback,rounded corners]
        (Check) -|
        (Act); 
    \draw[stream-tuple,rounded corners]
        (Act) |-
        (Predict); 
    \node [below of = Check] 
        (Legend) 
        { \usebox{\legendbox}
        };
    \end{tikzpicture}
        \caption{Assurance Monitor Agent Network With One Fusion Agent}
    \label{fig:network-of-agents}
\end{figure}
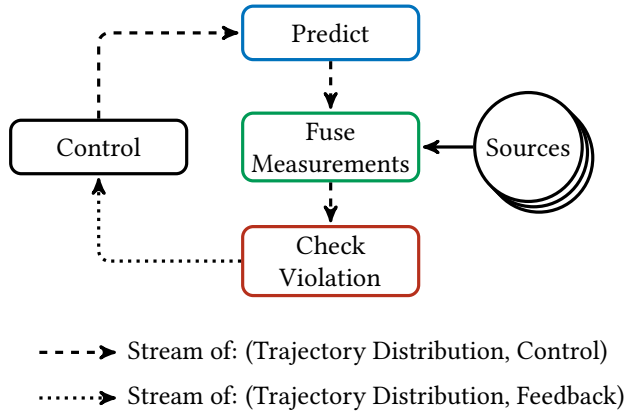

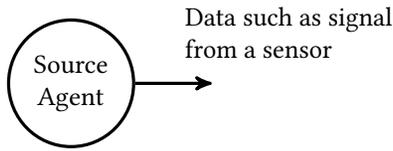
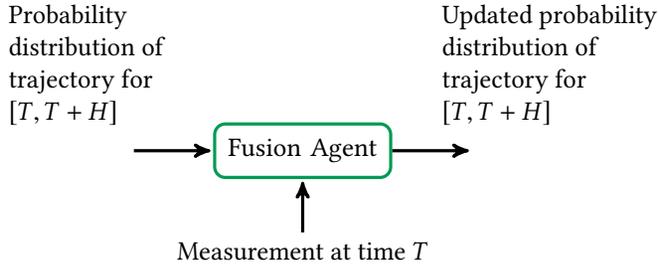
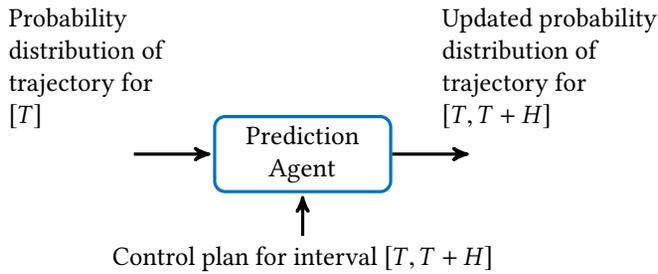
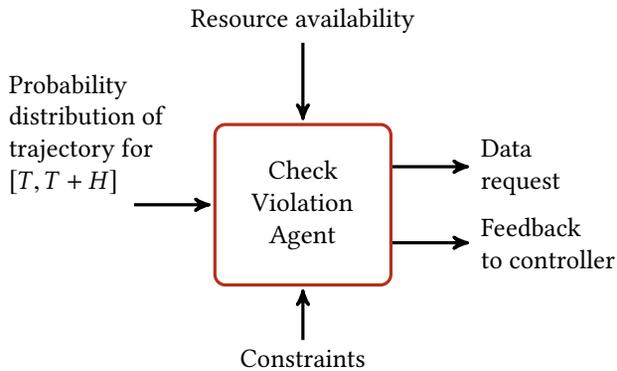
\begin{figure}

    \begin{subfigure}[b]{\linewidth}
        \centering
        \begin{tikzpicture}[node distance=1cm, auto,]
        \node [source-agent,align=center] (Source) {Source\\Agent};
        \node [right = 3em of Source] 
            (Sink) {};
        \draw[stream] 
            (Source) --
            node[above right,yshift=2mm,align=left] 
            { Data such as signal \\
              from a sensor
            }
            (Sink);
        \end{tikzpicture}
            \caption{Agent that Acquires Data}
        \label{fig:agent-source}
    \end{subfigure}
    
    \vspace{2.0em}
    
    \begin{subfigure}[b]{\linewidth}
        \centering
        \begin{tikzpicture}[node distance=1cm]
        \node [fusion-agent] (Fusion) {Fusion Agent};
        \node [left = 3em of Fusion] 
            (Source) {};
        \node [right = 3em of Fusion] 
            (Sink) {};
        \node [below = 2em of Fusion] 
            (Measurement) {Measurement at time $T$};
        \draw[stream] 
            (Source) --
            node[above left,yshift=2mm,align=left] 
            { Probability \\ 
              distribution of \\ 
              trajectory for \\ 
              $[T, T+H]$
            }
            (Fusion);
        \draw[stream] 
            (Fusion) --
            node[above right,yshift=2mm,align=left] 
            { Updated probability \\ 
              distribution of \\
              trajectory for \\
              $[T, T+H]$
            }
            (Sink);
        \draw[stream] 
            (Measurement) --
            (Fusion);
        \end{tikzpicture}
        \caption{Agent that Fuses Measurements}
        \label{fig:agent-fusion}
    \end{subfigure}
    
    \vspace{2.0em}
    
    \begin{subfigure}[b]{\linewidth}
        \centering
        \begin{tikzpicture}[node distance=1cm]
        \node [prediction-agent] (Prediction) {Prediction Agent};
        \node [left = 3em of Fusion] 
            (Source) {};
        \node [right = 3em of Fusion] 
            (Sink) {};
        \node [below = 2em of Fusion] 
            (Control) 
            { Control plan for interval $[T, T+H]$ 
            };
        \draw[stream] 
            (Source) --
            node[above left,yshift=2mm,align=left] 
            { Probability \\ 
              distribution of \\ 
              trajectory for \\ 
              $[T]$
            }
            (Prediction);
        \draw[stream] 
            (Prediction) --
            node[above right,yshift=2mm,align=left] 
            { Updated probability \\ 
              distribution of \\
              trajectory for \\
              $[T, T+H]$
            }
            (Sink);
        \draw[stream] 
            (Control) --
            node[below right,align=left] 
            {}
            (Prediction);
        \end{tikzpicture}
            \caption{Agent that Fuses Control Signals}
        \label{fig:agent-prediction}
    \end{subfigure}
    
    \vspace{2.0em}
    
    \begin{subfigure}[b]{\linewidth}
        \centering
        \begin{tikzpicture}[node distance=1cm]
        \node [check-agent,minimum height=6em] (Check) {Check Violation Agent};
        \node [above = 3em of Check] 
            (Resources) {Resource availability};
        \node [left = 3em of Check] 
            (Source) {};
        \node [ right = 3em of Check.east
              , anchor = west
              , yshift = 5mm
              , align = left ] 
            (Sink-Get-More-Data) 
            { Data \\ request
            };
        \node [ right = 3em of Check.east
              , anchor = west
              , yshift = -5mm
              , align = left ] 
            (Sink-Take-Action) 
            { Feedback \\ to controller
            };
        \node [below = 2em of Check] 
            (Constraints) {Constraints};
        \draw[stream] 
            (Resources) --
            (Check);
        \draw[stream] 
            (Source) --
            node[above left,align=left] 
            { Probability \\ 
              distribution of \\ 
              trajectory for \\ 
              $[T, T+H]$
            }
            (Check);
        \draw[stream] 
            ([yshift= 5mm]Check.east) to
            (Sink-Get-More-Data);
        \draw[stream] 
            ([yshift=-5mm]Check.east) to
            (Sink-Take-Action);
        \draw[stream] 
            (Constraints) --
            (Check);
        \end{tikzpicture}
        \caption{
            Agent that Predicts Assurance Violations
        }
        \label{fig:agent-check-violation}
    \end{subfigure}
    \caption{
        Assurance Monitor Agent Types
    }
    \label{fig:agent-types}
\end{figure}

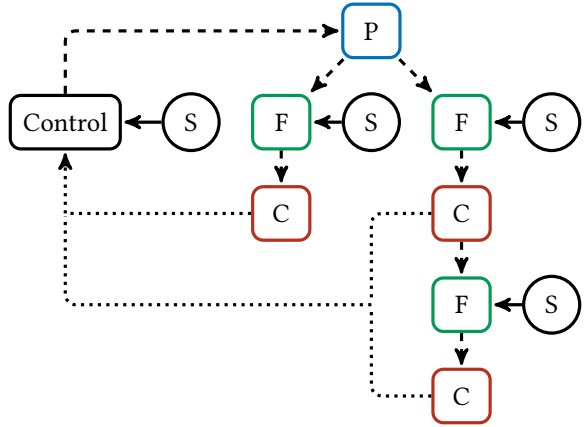
\begin{figure}
    \tikzset{
        mini-agent/.style={
                agent,
                very thick,
                text width=1.5em
        },
        mini-fusion-agent/.style={
                mini-agent,
                draw = ForestGreen
        },
        mini-prediction-agent/.style={
                mini-agent,
                draw = RoyalBlue
        },
        mini-check-agent/.style={
                mini-agent,
                draw = BrickRed
        },
        mini-source-agent/.style={
                mini-agent,
                circle,
                text width=1em
        },
        stream-tuple/.style={
                ->,
                very thick,
                shorten >=1pt,
                dashed
        },
        stream-feedback/.style={
                ->,
                very thick,
                shorten >=1pt,
                dotted
        },
        stream-feedback-noarrow/.style={
                very thick,
                shorten >=1pt,
                dotted
        }
    }
    \centering
    \begin{tikzpicture}[node distance=1.2cm]
    \node [mini-prediction-agent] 
        (Predict) 
        {P};
    \node [ mini-source-agent
          , below of = Predict
          ]
        (Source2) 
        {S};
    \node [ mini-fusion-agent
          , right of = Source2
          ] 
        (Fusion1) 
        {F};
    \node [ mini-source-agent
          , right of = Fusion1
          ]
        (Source1) 
        {S};
    \node [ mini-fusion-agent
          , left of = Source2
          ] 
        (Fusion2) 
        {F};
    \node [ mini-check-agent
          , below of = Fusion1
          ]
        (Check1) 
        {C};
    \node [ mini-check-agent
          , below of = Fusion2
          ]
        (Check2) 
        {C};
    \node [ mini-fusion-agent
          , below of = Check1
          ] 
        (Fusion3) 
        {F};
    \node [ mini-source-agent
          , right of = Fusion3
          ]
        (Source3) 
        {S};
    \node [ mini-check-agent
          , below of = Fusion3
          ]
        (Check3) 
        {C};
    \node [ mini-source-agent
          , left of = Fusion2
          ]
        (Source4) 
        {S};
    \node [ mini-agent
          , left = 1.5 em of Source4
          , text width = 3.5em
          , align=center
          ]
        (Act) 
        { Control
        };
    \coordinate [below of = Act]
        (join-below-act) {};
    \coordinate [left of = Fusion3] 
        (join-below-source2) {};
    \draw[stream-tuple] 
        (Predict) -- (Fusion1);
    \draw[stream-tuple] 
        (Predict) -- (Fusion2);
    \draw[stream] 
        (Source1) -- (Fusion1);
    \draw[stream] 
        (Source2) -- (Fusion2);
    \draw[stream-tuple] 
        (Fusion1) -- (Check1);
    \draw[stream-tuple] 
        (Fusion2) -- (Check2);
    \draw[stream-tuple] 
        (Check1)  -- (Fusion3);
    \draw[stream] 
        (Source3) -- (Fusion3);
    \draw[stream-tuple] 
        (Fusion3) -- (Check3);
    \draw[stream] 
        (Source4) -- (Act);
    \draw[stream-feedback-noarrow,rounded corners]
        (Check2) -- (join-below-act);
    \draw[stream-feedback,rounded corners]
        (join-below-act) -- 
        (Act);
    \draw[stream-feedback-noarrow,rounded corners]
        (Check3) -| 
        (join-below-source2);
    \draw[stream-feedback-noarrow,rounded corners]
        (Check1) -| 
        (join-below-source2);
    \draw[stream-feedback-noarrow,rounded corners]
        (join-below-source2) -| 
        (join-below-act);
    \draw[stream-tuple,rounded corners]
        (Act)     |- (Predict);
    \end{tikzpicture}
        \caption{Assurance Monitor Agent Network With Agents Composed Both Sequentially and in Parallel}
    \label{fig:complex-network-of-agents}
\end{figure}

An assurance monitor pattern is a network of different types of
agents (Figure~\ref{fig:network-of-agents}). Each agent is a smaller pattern in itself.  A \textit{data source} agent (Figure~\ref{fig:agent-source}) acquires data, for example from a sensor; this data is
acquired when the agent is requested to do so by a decision maker
agent or by a system clock.  A data \textit{fusion} agent (Figure~\ref{fig:agent-fusion}) reads an
estimate of the probability distribution over system states and data
from a set of sources and outputs a new estimate of the distribution.
A \textit{prediction} agent (Figure~\ref{fig:agent-prediction}) reads an estimate of the probability distribution over system states and a control signal and outputs a new estimate of the distribution.
A \textit{check violation} agent (Figure~\ref{fig:agent-check-violation}) computes the probability of an
assurance violation event. The input to a check-violation agent is a
probability distribution on the trajectory; it outputs either:
\begin{enumerate}
\item an alert signaling the need for a course correction, or
\item a request to get additional data from a set of data sources, or
\item a continue message indicating that the current plan is safe.
\end{enumerate}
\textit{Computation agents} look up data repositories or carry out
computations to add value to the data. For example, a computation
agent may integrate stored historical data with raw sensor data.

The agent network in Figure~\ref{fig:network-of-agents} contains a single fusion agent. More complex networks can be constructed, where the signals of source agents are composed sequentially (as we will see in Section~\ref{sec:drone-navigation}) or in parallel, where agents execute concurrently. The network in Figure~\ref{fig:complex-network-of-agents} shows a network where agents are composed in both ways.

\subsection{Consequences}

Autonomous systems, particularly those whose behavior relies on components with partially unknown behavior such as those based on learned models, can incorporate an assurance monitor to retain some of the operational safety of manually controlled or manually programmed systems. The assurance monitor can aid in conserving scarce resources, by only activating system functions that consume these resources when necessary.
The composition of fusion agents that propagate a stream of successive state estimates can increase the accuracy of the overall system beyond the accuracy of any individual agent, and compensate for several kinds of inaccuracies, including drift, noise and approximation.

These benefits come at the cost of runtime overhead of the computations that are required for assurance monitoring, which may be considerable. As with most predictive systems, the detection of future assurance violation events can yield false positives. Under unfavorable conditions, or if the violation criteria are too conservative, the resulting actions may cause unnecessary remedial actions or instability in a system that would otherwise remain both safe and stable.\\

Next, we describe applications constructed by networks of these agents.

\section{Example Application: Resource-Constrained Autonomous Drone Navigation}
\label{sec:drone-navigation}

To help understand what instances of the pattern can look like, we begin by reviewing an example that exhibits several characteristics that make the pattern useful.
The application is a drone, that is required to travel over a region while staying away from no-fly zones. The drone operates autonomously using an on-board control system. Some recently proposed control systems for autonomous drones are entirely data-driven, meaning that the behavior of the drone is learned from a training set of sensor inputs~\cite{smolyanskiy2017toward}, which may even be captured on a different system~\cite{loquercio2018dronet}. The potential for autonomy in this approach is promising, but the learned behavior is generally not combined with manual specifications, such as safety properties, meaning that the overall system may be unsafe if exposed to conditions not covered by the training set. An assurance monitor can serve as a bridge between these two worlds, yielding a system that is both autonomous and avoids assurance violations, for example by switching to a safe controller when the learned controller is predicted to take the system out of safety.

In our example, the drone has accelerometers and gyroscopes that enable
it to estimate its 3-D acceleration. It takes photographs of
the terrain periodically, and the drone may take additional
photographs to improve location estimation. The drone may also
activate other mechanisms, such as GPS, to improve its estimate. The
use of these mechanisms consumes power and other resources, and so
they must be used judiciously.

The state of the drone is given by its (3-D) acceleration, velocity
and location. A network of agents computes a probability distribution
of the drone's current state at a point in time. The distribution at
time $t=0$ is given. The drone's trajectory over a time interval is
its state at each point in the interval. The agents compute the
distribution of trajectories to detect events that trigger actions
such as ``get more data'' or ``change course.''

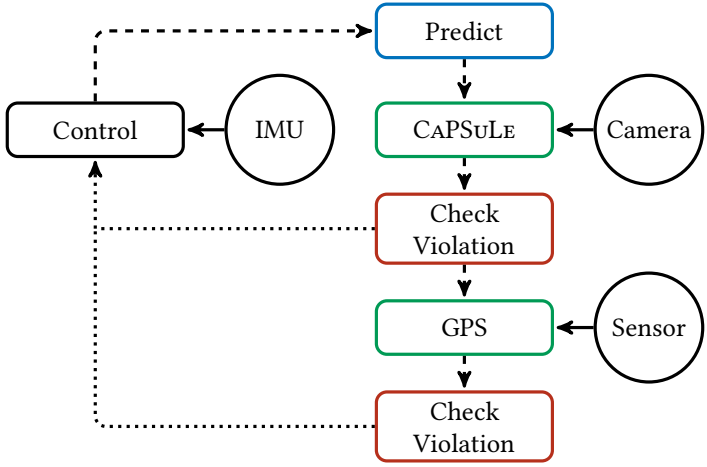
\begin{figure}
    \tikzset{
        stream-tuple/.style={
                ->,
                very thick,
                shorten >=1pt,
                dashed
        },
        stream-feedback/.style={
                ->,
                very thick,
                shorten >=1pt,
                dotted
        },
        stream-feedback-noarrow/.style={
                very thick,
                shorten >=1pt,
                dotted
        }
    }
    \centering
    \begin{tikzpicture}[node distance=1.3cm]
    \node [prediction-agent] 
        (Predict) 
        {Predict};
    \node [ fusion-agent
          , below of = Predict
          ] 
        (Capsule) 
        {\capsule};
    \node [ source-agent
          , right = 1.5em of Capsule
          ]
        (Camera) 
        {Camera};
    \node [ check-agent
          , below of = Capsule
          ]
        (Check-Capsule) 
        {Check Violation};
    \node [ fusion-agent
          , below of = Check-Capsule
          ]
        (GPS) 
        {GPS};
    \node [ source-agent
          , right = 1.5em of GPS
          ]
        (Sensor) 
        {Sensor};
    \node [ check-agent
          , below of = GPS
          ]
        (Check-GPS) 
        {Check Violation};
    \node [ source-agent
          , left = 1.5em of Capsule
          , align=center
          ]
        (IMU) 
        { IMU
        };
    \node [ agent
          , left = 1.5em of IMU
          , align=center
          ]
        (Act) 
        { Control
        };
    \coordinate [below of = Act] (join) {};
    \draw[stream-tuple] 
        (Predict) --
        (Capsule);
    \draw[stream] 
        (Camera) --
        (Capsule);
    \draw[stream-tuple] 
        (Capsule) --
        (Check-Capsule);
    \draw[stream-tuple] 
        (Check-Capsule) --
        (GPS);
    \draw[stream] 
        (Sensor) --
        (GPS);
    \draw[stream-tuple] 
        (GPS) --
        (Check-GPS);
    \draw[stream-feedback-noarrow]
        (Check-Capsule) --
        (join); 
    \draw[stream-feedback-noarrow,rounded corners]
        (Check-GPS) -|
        (join); 
    \draw[stream-feedback]
        (join) --
        (Act); 
    \draw[stream-tuple,rounded corners]
        (Act) |-
        (Predict); 
    \draw[stream] 
        (IMU) --
        (Act);
    \end{tikzpicture}
        \caption{Drone Navigation Assurance Monitor Agent Network}
    \label{fig:navigation-network-of-agents}
\end{figure}

Figure~\ref{fig:navigation-network-of-agents} shows the network of 
agents that make up the drone navigation assurance monitor. 
Next, we describe the behavior of its key agents.

\subsection{Source Agents}

The drone's inertial measurement unit (IMU, which consists of an 
accelerometer and/or gyroscope) are read by a source agent
which produces a continuous stream of acceleration measurements.  At
time $T$ a prediction agent computes a probability distribution of the
drone's trajectory in an interval $[T, T+H]$, where $H$ is a time
horizon, given a probability distribution of the drone's state at time
$T$ and the drone's planned acceleration in the interval $[T,
T+H]$. The true acceleration is the planned acceleration plus noise.

\subsection{Fusion Agents}

Many fusion agents use the Bayesian update formula, and for
completeness, we give the equation. In this example, probabilities and
time steps are assumed to be discrete.  
Let $a(t)$ be the planned acceleration at each point in the interval
$[T, T+H]$, and let $a$ be the vector of accelerations over the
interval. We do not discuss how $a$ is calculated. Let random variable
$e(t)$ be the acceleration noise at time $t$. The noise may be correlated
with the state (location, velocity and acceleration) of the drone. Let
$g(z,a)$ be the probability that the trajectory is $z$ in the interval
$[T, T+H]$ given the distribution of drone states at time $T$, the
noise distribution $e$, and accelerations $a$.
\begin{equation} \label{eqn:trajectory-probability}
    g(z,a) \triangleq \sum_{s, e} p(trajectory = z | a, e, s)*p(state_{t}=s)*p(noise=e)
\end{equation}

\noindent Since fusion agents update a probability distribution based on some additional data, the output of such an agent is a probability distribution, which can be used as the input to another fusion agent. Figure~\ref{fig:navigation-network-of-agents} shows an example of such a composition of fusion agents.

\subsection{Fusion Agent: \capsule{}}
\label{sec:fusion-agent-capsule}

The \capsule{} fusion agent implements a positioning system that approximates the location of the drone through an image-based matching algorithm~\cite{moon2016capsule} that utilizes the camera of that device.
The location is approximated by matching the image, captured by the on-board camera, against a database of images stored on-board.

Figure~\ref{fig:terrain_simulation_capsule} shows that the raw location distribution reported by \capsule{} can be both uncertain and inaccurate under adverse conditions. This is a result of the input image that is fed to \capsule{}, corresponding to the area under the yellow rectangle, which is mostly cloud-covered water and lacks the distinct features that are needed for accurate matching. The uncertainty and error are mitigated by the \capsule{} fusion agent in the same simulation step, by fusing the \capsule{} distribution with the previous location distribution, updated with respect to the previous control action. 
Given the location distribution $L_{T'}$ at the previous time $T'$, updated with respect to the control signal for the time interval $(T',T]$, and given the location distribution $C_T$ obtained from \capsule{} at the current time $T$, the location distribution $L_T$ for $T$ is computed as follows:
\begin{equation}
L_T^{x,y} =
    L_{T'}^{x,y}\cdot\hspace{-2em}
        \sum_{(x', y') \in dim(C_T)}\hspace{-2em}
            (C^{x', y'}_T \cdot K_{x,y}^{x', y'})
\end{equation}
\noindent where $L^{x,y}$ denotes the probability density at location $(x,y)$ in the distribution $L$, $dim(C_T)$ is the set of coordinates of $C_T$, and $K_{x,y}^{x', y'}$ is the probability that an image taken at $(x,y)$ matches location $(x', y')$. The map $K$ is estimated by feeding each image in the \capsule{} image database as a query to \capsule{}, thus obtaining a distribution for the corresponding location.

The resulting distribution is both more precise, and more accurate, as we can see in Figure~\ref{fig:terrain_simulation_fused}. The simulation suggests that accuracy of camera-based navigation can be improved substantially by taking into account other information that is available on-board a drone during a flight, even when the drone is controlled autonomously. We note that, even though Figure~\ref{fig:terrain_simulation_fused} shows fusion improving the \capsule{} location estimate, Figure~\ref{fig:terrain_simulation_capsule} is a particularly imprecise example, and there are situations when the fusion will instead correct the location estimate obtained from the control signal. This is because the latter is subject to drift, due to external forces acting on the drone, aside from the actuation that the controller dictates.

A computation agent matches photos taken by the drone with the database to compute the conditional probability $p(s | q)$ that the drone is at location $s$ given photo $q$. 
This probability is computed by \capsule{}~\cite{moon2016capsule} in the experiments reported in Section~\ref{sec:experiment}; we note that any program with a Python interface can be inserted easily into the the computation agent to replace \capsule{} with another localization algorithm.

Several factors contribute to uncertainty in the location distribution output by \capsule{}. If the query image depicts an area that does not possess unique features, then the resulting distribution will be spread out across the map, reflecting the locations where such features are present.
However, physical constraints, such as the previous location of the drone and the plan that was used to control the drone since then, dictate that some of those possible locations are unrealistic. This intuition is realized in the updating scheme implemented by the fusion agent, whereby the previous location distribution, updated with respect to the control signal (by the Predict agent) is fused with the current \capsule{} location estimate.

\subsection{Fusion Agent: GPS}

The GPS fusion agent reads a signal from its sensor, which is assumed to coincide with the ground truth location with a high probability $p_{\textup{GPS}}$.
Thus, compared to the inexact information obtained from the \capsule{} agent, the GPS signal plays the role of an oracle for the location, but does so at the cost of consuming additional resources.
The error is assumed to be uniformly distributed over its eight neighboring locations, which are each assigned a probability of $\frac{1 - p_{\textup{GPS}}}{8}$.

\subsection{Check Violation Agents}

A drone is a resource-constrained system. Aside from navigation, 
its on-board energy resources are used for propulsion, to power sensors 
and communications equipment, and to perform computation needed for the
drone's mission. Under these circumstances, it is of interest to use 
cheaper alternatives to resource-intensive processes, such as positioning,
whenever possible. Yet, when the risk of some harm is deemed sufficiently
high, resources should be used to mitigate the risk. Managing this tradeoff 
is the job of a check-violation agent.

A check-violation agent uses $g(z)$ (defined in Equation~\ref{eqn:trajectory-probability}) to compute the probability 
that the drone will enter a no-fly zone in $[T, T+H]$. This agent also
checks the availability of resources required to acquire and process
additional sensor data. If the probability of violating constraints is
low then the agent generates a continue signal and the drone continues
its planned acceleration. Otherwise, if sufficient resources (energy,
computational power) are available, the agent requests that the system
acquire location information from the GPS sensor to improve the drone's
location estimate. If the probability of violating constraints is
unacceptable and insufficient resources are available, the agent
generates an event signaling that a correction is required. 

The optimal strategy for the check-violation agent is complex and the
experiments use a simpler heuristic. If the probability of violation
is low then the agent outputs a continue message regardless of the
availability of resources. 
Else, if the amounts of resources available exceed a threshold then the 
agent outputs a request to get additional data. This data should produce 
better estimates (in some cases, images may be taken for goals other than 
navigation; in these cases the agent requests use of the images). 
Finally, if the agent determines that violation is likely and that resources 
are insufficient, then the agent outputs a message instructing the drone
to take corrective action.

Figure~\ref{fig:terrain_simulation_check_violation} illustrates the predictions (blue rectangles) generated by the Check Violation agent, and resulting corrective control action (red line). The assurance monitor makes decisions based on predictions over a time horizon $H$ into the future. Based on $H$, and the location distribution $L_T$ at the current time $T$ obtained from the corresponding fusion agent, a sequence $F_{[T,T+H]}$ of location distributions for the time interval $[T,T+H]$ is computed from $L_T$ by propagating the control plan $a(t)$, augmented by an estimate of future perturbations $b(t)$, up to time $T+H$.
The trajectory map $F_{[T,T+H]}$ is a sequence of location distributions $F_0,F_1,\ldots,F_{H-1}$, computed using the distribution propagation map $G$ as follows:
\begin{eqnarray*}
F_0 & = & L_T \\
F_n & = & G(F_{n-1}) \\
G(L)^{x,y} & = & L^{d(x,y)} + e(L,x,y)\\
d(x,y) & = & (x,y) - a(t) - b(t)
\end{eqnarray*}
\noindent where $a(t)$ and $b(t)$ are time-indexed sequences of spatial displacements, and $e(L,x,y)$ is a noise term by which the probability at location $(x,y)$ dissipates into its neighboring locations with a given probability.

\section{Experiments in Assurance Monitoring}
\label{sec:experiment}

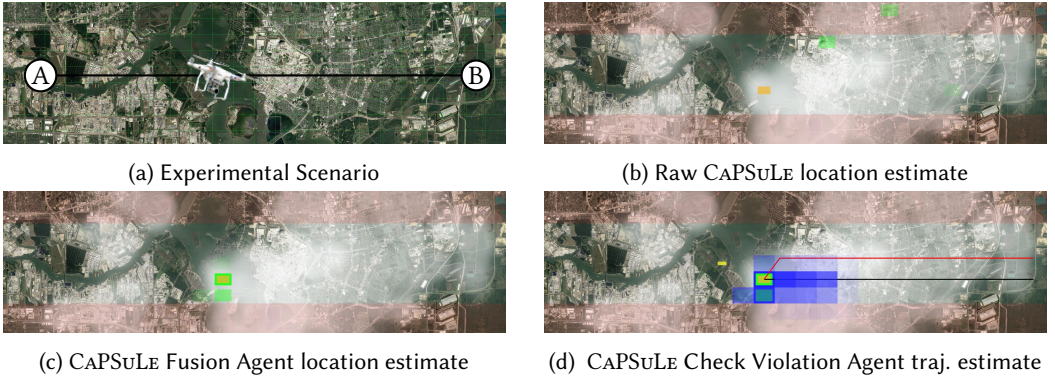
\begin{figure*}
    \centering
    \begin{subfigure}[b]{0.48\linewidth}
        \begin{tikzpicture}
            \node[anchor=south west, inner sep=0] at (0,0) {
                \adjincludegraphics[width=\linewidth,trim={{0} {0.26\height} {0} {0.3\height}},clip]{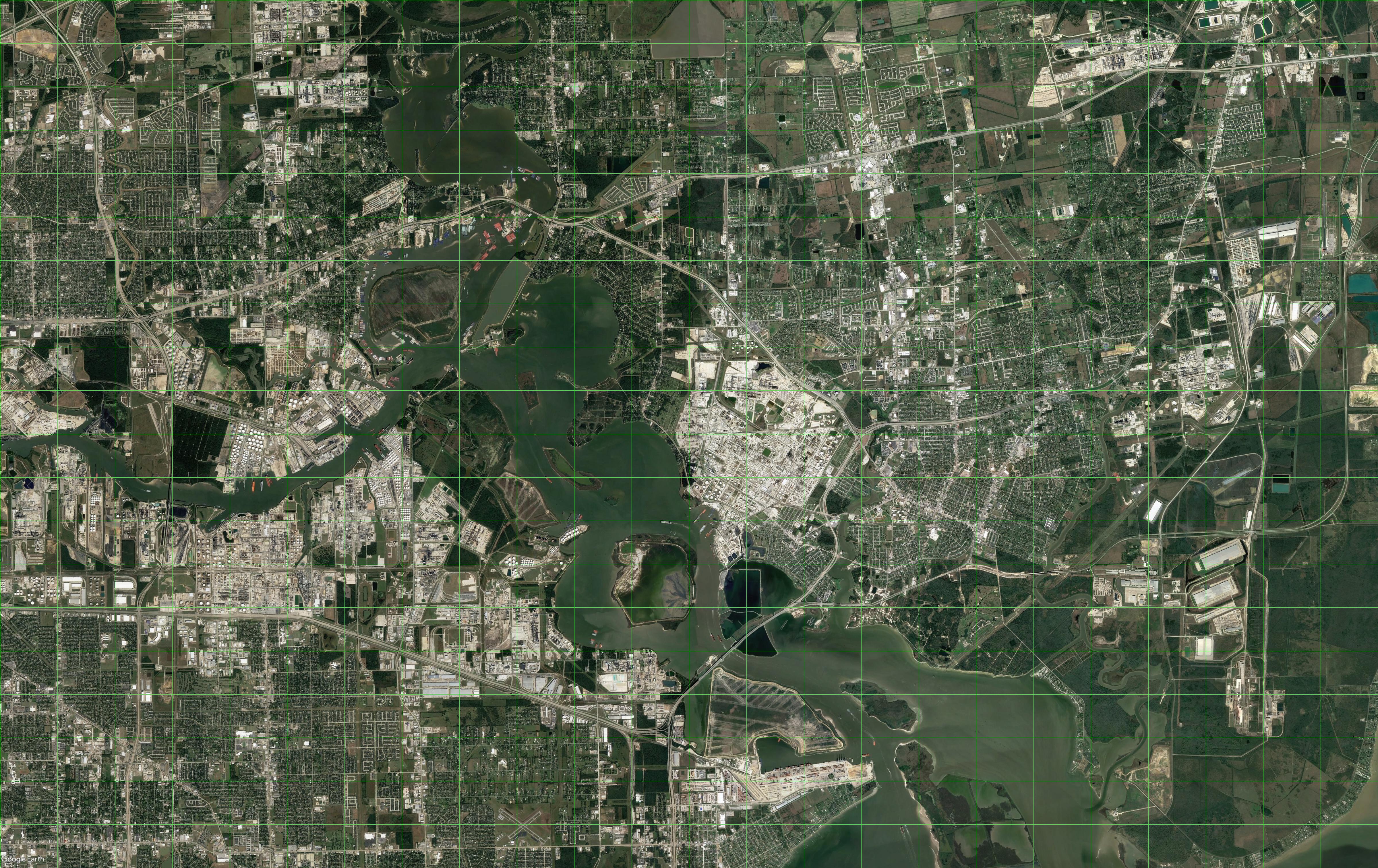}
            };
            \draw[black,thick] (0.5,0.9) -- (6.3,0.9);
            
            \node[anchor=south west, inner sep=0,rotate=-20] at (2.3,0.75) {
                \includegraphics[width=0.15\linewidth]{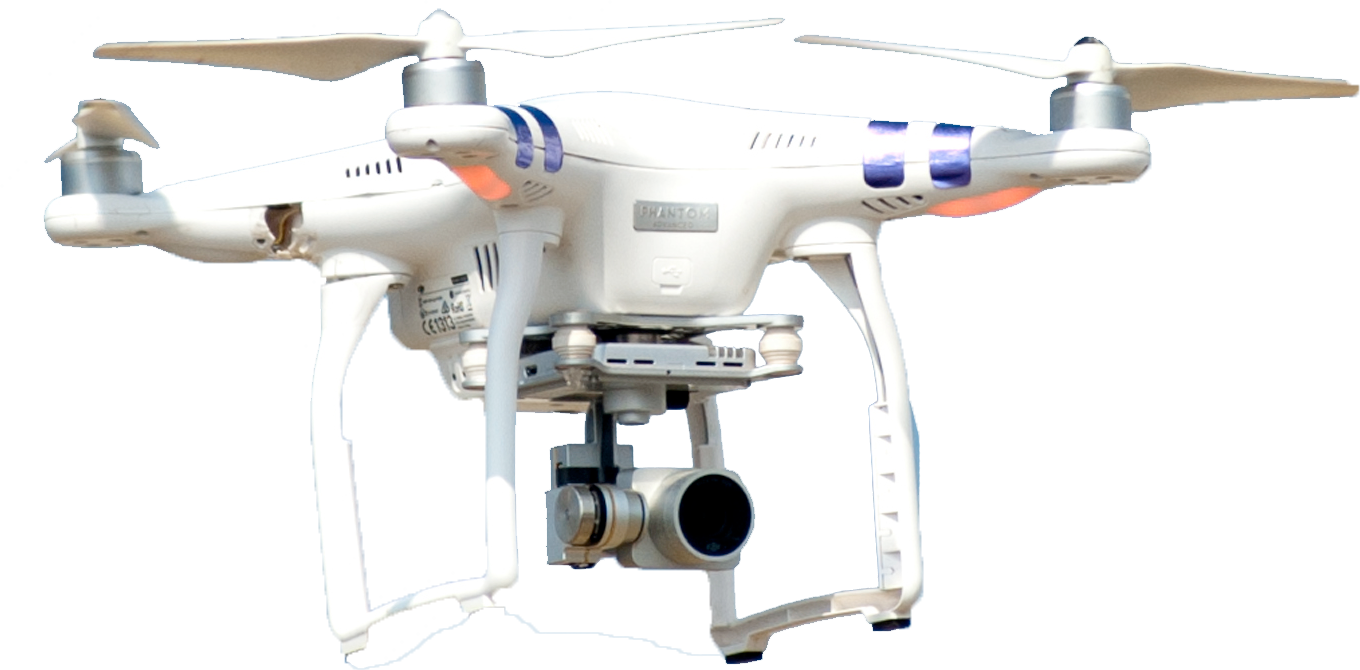}
            };
            \node[circle,fill=white,draw=black,thick,inner sep=0.3mm] 
                (A) at (0.5,0.9) {A};
            \node[circle,fill=white,draw=black,thick,inner sep=0.3mm] 
                (B) at (6.3,0.9) {B};
        \end{tikzpicture}
        \caption{Experimental Scenario}
        \label{fig:terrain_simulation_overview}
    \end{subfigure}\hspace{1.5em}%
    \begin{subfigure}[b]{0.48\linewidth}
        \adjincludegraphics[width=\linewidth,trim={{0} {0.26\height} {0} {0.3\height}},clip]{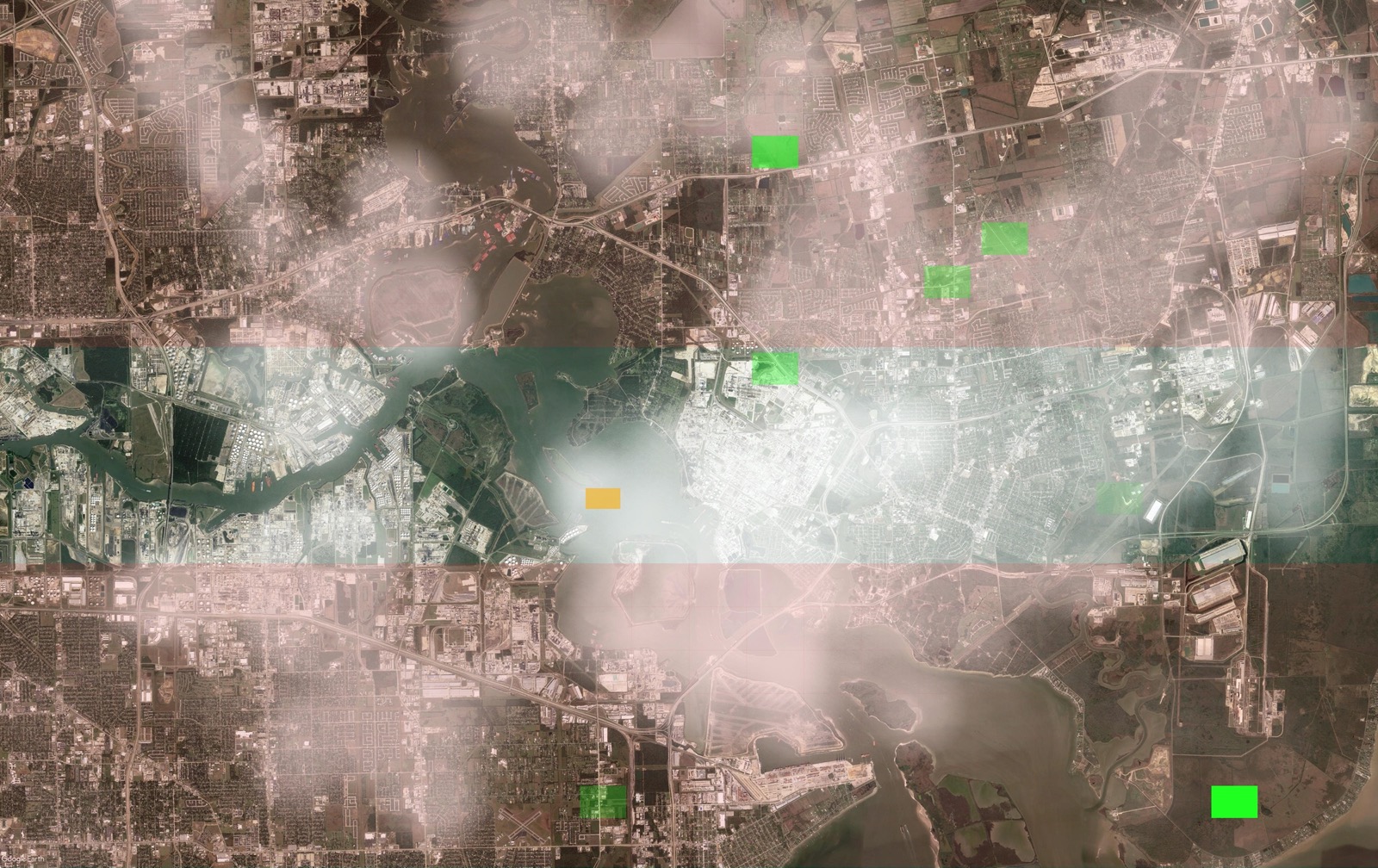}
        \caption{Raw \capsule{} location estimate}
        \label{fig:terrain_simulation_capsule}
    \end{subfigure}
    
    \begin{subfigure}[b]{0.48\linewidth}
        \adjincludegraphics[width=\linewidth,trim={{0} {0.26\height} {0} {0.3\height}},clip]{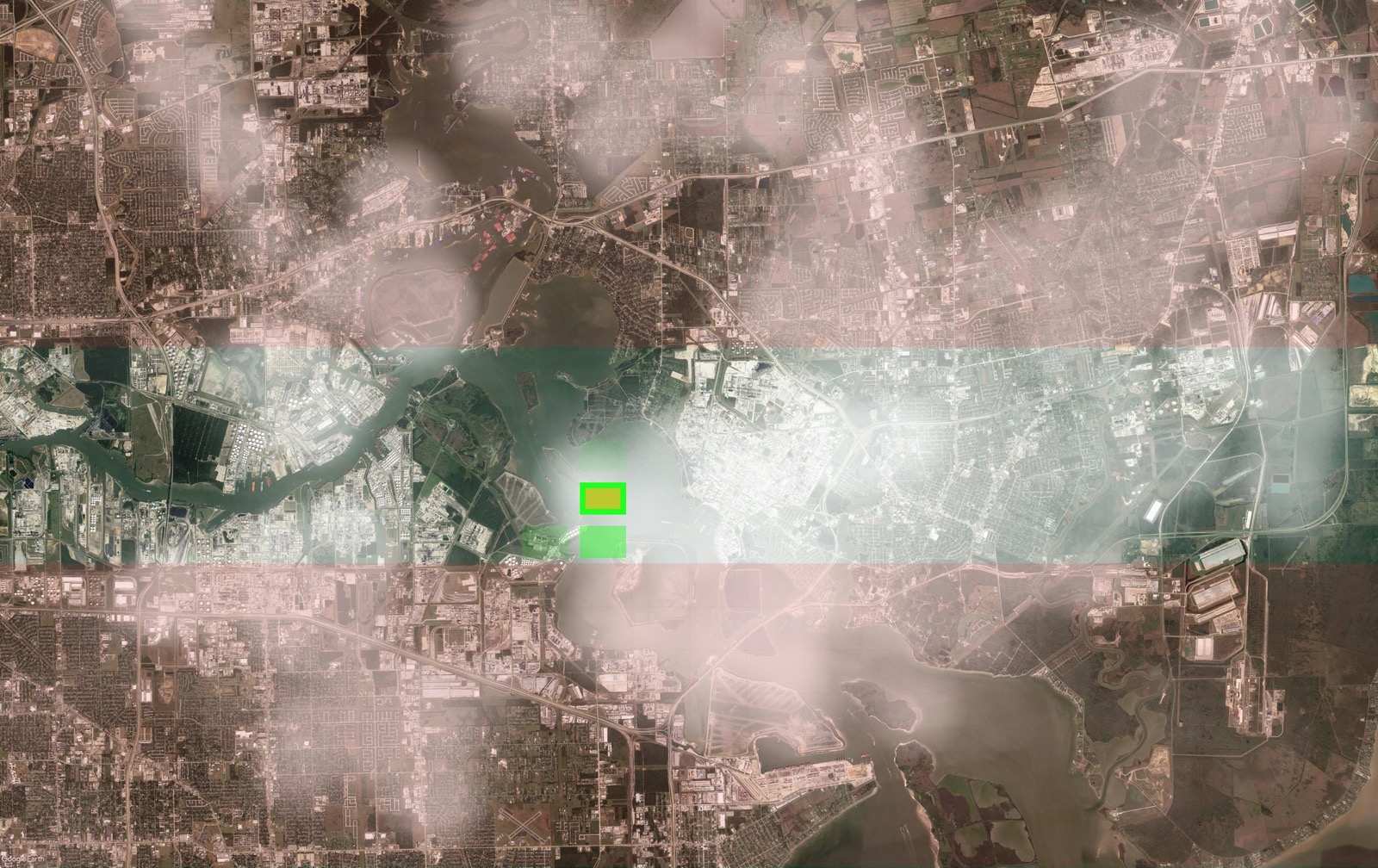}
        \caption{\capsule{} Fusion Agent location estimate}
        \label{fig:terrain_simulation_fused}
    \end{subfigure}\hspace{1.5em}%
    \begin{subfigure}[b]{0.48\linewidth}
        \adjincludegraphics[width=\linewidth,trim={{0} {0.26\height} {0} {0.3\height}},clip]{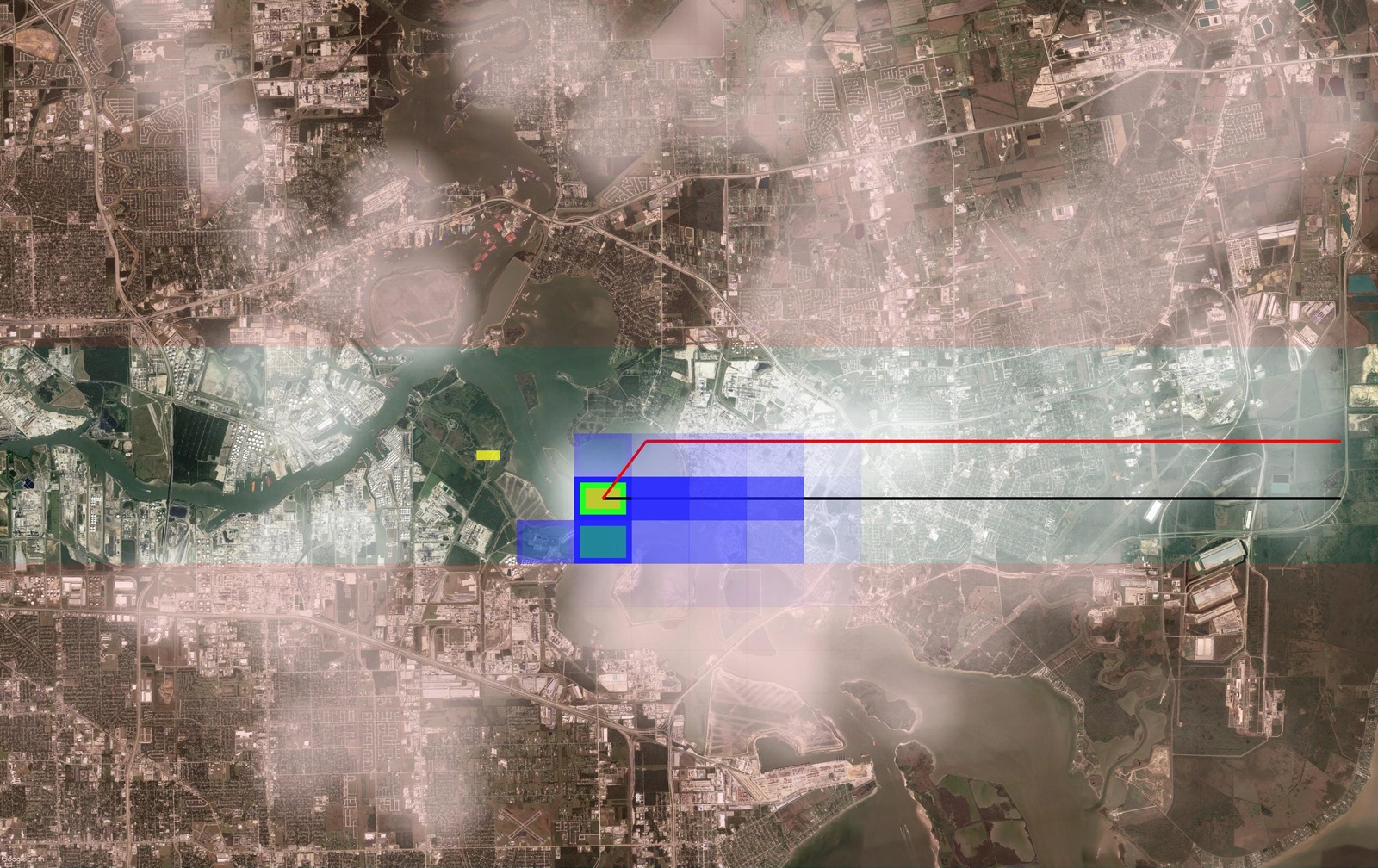}
        \caption{\label{fig:terrain_simulation_check_violation} \capsule{} Check Violation Agent traj. estimate}
    \end{subfigure}
    \caption{
        Drone Navigation Simulation.
        Red areas represent the no-fly zone. A yellow rectangle represents the originally planned position. A yellow rectangle represents the ground truth, which is determined by both control and perturbations. Green rectangles represent the current location distribution. Blue rectangles represent planned trajectory distribution used by the assurance monitor to estimate probability of entering the no-fly zone.
    }
    \label{fig:terrain_simulation}
\end{figure*}

To demonstrate the behavior of an assurance monitor, we review a simulation of the drone navigation application described in the previous section. 
In this simulation, the drone should follow a straight-line trajectory between two points $A$ and $B$, at either side of the known terrain illustrated in Figure~\ref{fig:terrain_simulation_overview}. This terrain is represented as a grid of images in the database, constructed by partitioning a high-resolution aerial photograph of the terrain into $20 \times 24$ tiles. Each tile is associated with the location at its center point.

While the photograph is taken in good weather, the simulation assumes that the drone is flying in less favorable conditions. Thus, in the images that are available for navigation, parts of the terrain is obscured by clouds, making the camera-based navigation task harder, and the location distribution reported by \capsule{} more uncertain.

The controller's planned actions are represented by a list of velocities for the remainder of the flight. During the flight, wind and actuation inaccuracies take the drone off its planned trajectory. This is simulated by a random perturbation that is added to the control velocities at every step of the simulation.

Locations more than two tiles away from the planned trajectory are considered a \emph{no-fly zone}, colored  in red. The job of the assurance monitor is to predict that the drone may enter this zone -- as a result of incorrect control, the past history of perturbations -- and initiate corrective control. To avoid unnecessary avoidance maneuvers due to uncertainty about the location of the drone, the assurance monitor implements a policy that activates the GPS when the remaining resources exceed a threshold, which is a function of the remaining energy resources and flight duration.

\subsection{Assurance}

The opacity of the blue rectangles in Figure~\ref{fig:terrain_simulation_check_violation} reflect the densities of locations in $F_{[T,T+H]}$. The probability of entering the no-fly zone is defined as the greatest density among all locations in $F_{[T,T+H]}$ that are also in the no-fly zone. Figure~\ref{fig:terrain_simulation_check_violation} shows a step in the simulation where this probability exceeds the $5\%$ threshold for an assurance violation, causing the system to generate a corrective control action. To take into account uncertainty about the current state of the system, the assurance monitor performs its predictions based on distributions. However, a single control action must be passed to the controller. Thus, the revised control plan (red line) is computed based on the mean location, which can be seen to take the drone to the target location B along a safer trajectory, further away from the no-fly zone, compared to the original control plan (black line).

Table~\ref{tab:decisions} shows an extract from an execution of the simulation that exhibits different signals produced by \capsule{} and GPS check violation (CV) agents.
Up to time 12, the \capsule{} fusion actor's location estimate suffices to yield a sequence of ``Continue'' decisions from the \capsule{} CV agent.
At time 13 the drone is taken off-course by wind. The probability of entering the no-fly zone remains under the $5\%$ threshold over the next two steps, which again yields ``Continue'' signals from the \capsule{} CV agent.
At time 15, wind again moves the drone. This time, the trajectory distribution indicates a high ($30.9\%$) probability of entering the no-fly zone. Because sufficient resources remain, the \capsule{} CV agent produces the ``More Data'' signal. The subsequent GPS reading is sufficiently precise that the probability of entering the no-fly zone falls back below the threshold, and the GPS CV agent produces a ``Continue'' signal.
However, at the next step, both the \capsule{} and GPS CV agents predict a high risk of assurance violation, and the GPS CV agent produces a ``Change'' signal to redirect the drone to a safe trajectory.
Over the next three steps, the drone moves away from the no-fly zone along the revised, safe trajectory. The trajectory estimates reflect the successively falling probability of entering the no-fly zone during these steps.

\begin{table}[htbp]
\caption{
    Drone navigation simulation steps that exhibit different
    check violation agent dections.
}
\centering
\newcommand{\checkViolationStep}[1]{\multirow{2}{*}{\adjincludegraphics[width=0.23\linewidth,trim={{0.425\width} {0.3\height} {0.05\width} {0.34\height}},clip]{images/scenario_table/check_violation_heat_map_#1.jpg}}}
\setlength\dashlinedash{0.2pt}
\setlength\dashlinegap{1.5pt}
\setlength\arrayrulewidth{0.3pt}
\newcommand{\hdottedline}{\hdashline\vspace{-0.9em}\\}
\begin{tabular}[t]{ 
    @{}
    l@{\hspace{0.73em}}
    l@{\hspace{0.73em}}
    l@{\hspace{0.73em}}
    l@{\hspace{0.73em}}
    l@{\hspace{0.73em}}
    l@{\hspace{0.73em}}
    @{\hspace{-0.75em}}
}
\toprule
Predicted trajectory & Time & Resources & Probability & CV Agent & Signal \\
\midrule
\checkViolationStep{12}
& 12 & 12 & 0.1\% & \capsule{} & Continue\\\
& & & & & \vspace{1.83em} \\
\hdottedline
\checkViolationStep{13}
& 13 & 12 & 1.9\% & \capsule{} & Continue\\\
& & & & & \vspace{1.83em} \\
\hdottedline
\checkViolationStep{14}
& 14 & 12 & 2.0\% & \capsule{} & Continue\\\
& & & & & \vspace{1.83em} \\
\hdottedline
\checkViolationStep{15}
& 15 & 12 & 30.9\% & \capsule{} & More Data \\
& & 11 & 1.9\% & GPS & Continue \vspace{1.83em}\\
\hdottedline
\checkViolationStep{16}
& 16 & 11 & 27.5\% & \capsule{} & More Data \\
& & 10 & 34.1\% & GPS & Change \vspace{1.83em}\\
\hdottedline
\checkViolationStep{17}
& 17 & 10 & 2.14\% & \capsule{} & Continue \\
& & & & & \vspace{1.83em} \\
\hdottedline
\checkViolationStep{18}
& 18 & 10 & 1.9\% & \capsule{} & Continue \\
& & & & & \vspace{1.83em} \\
\hdottedline
\checkViolationStep{19}
& 19 & 10 & 0.2\% & \capsule{} & Continue \\
& & & & & \vspace{1.7em} \\
\bottomrule
\end{tabular}
\label{tab:decisions}
\end{table}

The experiment described in this section uses assurance criteria based on the system's physical location. The assurance criterion could easily be extended to depend on more abstract states, such as temperature, current, power, latency or remaining resources, using the same pattern to avoid that it is violated.

\section{Example Application: Timestamp Sensors}
\label{sec:another-example}

The assurance monitoring pattern uses networks of agents including
source, fusion, prediction and check-violation agents
(Figure~\ref{fig:network-of-agents}). 
Tracking airplanes and taking appropriate action using Kalman filters
is an example of the assurance monitoring pattern that is very widely
used. The pattern is also used in finance where the constraints are on
deviation of the performance of a stock portfolio from benchmarks of
the broad stock exchange. In this section, we review an example from 
sensor networks. Our goal is limited to showing how the same pattern 
occurs in a very different context from that covered in 
Section~\ref{sec:drone-navigation}. 

Some event-based applications require the assurance that events are
timestamped accurately. For example, accurate timestamps are
necessary in applications that analyze mechanical or seismological
structures by computing relative accelerations of structural
components~\cite{clayton2012community}. In some cases, computers
attached to sensors -- such as those in utility closets in building
basements -- do not have access to GPS, and in these cases, timestamps
are obtained from local system clocks in the computer. The accuracy of timestamps
can be improved by using NTP services; however, the use of such
services requires communication and computation which can be scarce
resources on embedded computers that also have to perform other
tasks. 

A constraint for the application is that the time reported by the
local clock can deviate from the true time by no more than a specified
limit. The application is required to provide the assurance that this
constraint is never violated or is violated only with very low
probability.

Using the local clock is inexpensive compared with using the NTP
server, and so the application determines when to call NTP. In this
application, the resource that has to be managed is the amount of
bandwidth and computation used by the application whereas in drone
navigation the resources were battery power and computational load.

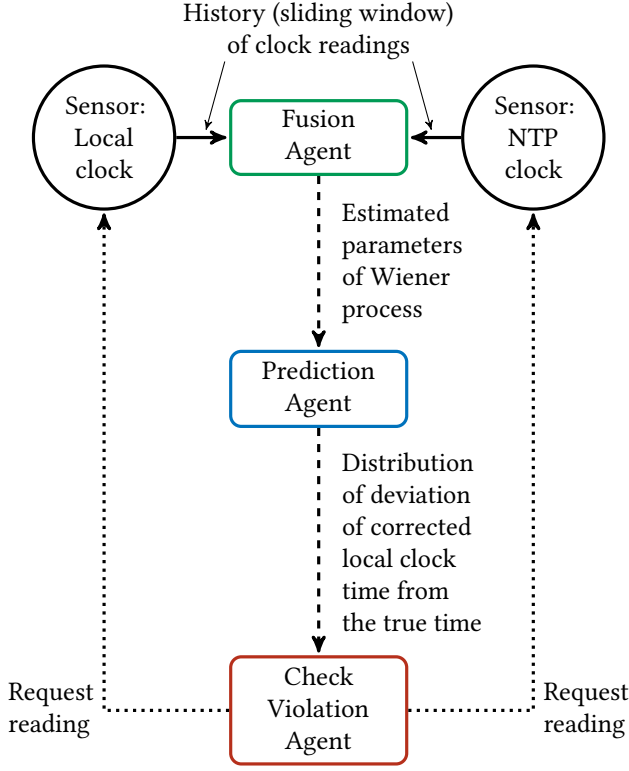
\begin{figure}
    \tikzset{
        stream-tuple/.style={
                ->,
                very thick,
                shorten >=1pt,
                dashed
        },
        stream-feedback/.style={
                ->,
                very thick,
                shorten >=1pt,
                dotted
        },
        stream-feedback-noarrow/.style={
                very thick,
                shorten >=1pt,
                dotted
        }
    }
    \centering
    \begin{tikzpicture}[node distance=1.3cm]
        \node[above = 1.5em of Fusion, align=center] 
        (Reading-Description)
        {History (sliding window)\\of clock readings};
        \draw[->,thin] ( 1.25,1.0) -- ( 1.5,0.1);
        \draw[->,thin] (-1.25,1.0) -- (-1.5,0.1);
        \node [ fusion-agent] 
            (Fusion) 
            {Fusion\\ Agent};
        \node [ prediction-agent
              , below = 6.5em of Fusion
              ] 
            (Predict) 
            {Prediction Agent};
        \node [ source-agent
              , left = 2.0em of Fusion
              ]
            (Sensor-local-clock) 
            {Sensor: Local clock};
        \node [ source-agent
              , right = 2.0em of Fusion
              ]
            (Sensor-ntp-clock) 
            {Sensor: NTP clock};
        \node [ check-agent
              , below = 8.5em of Predict
              ]
            (Check) 
            {Check \\ Violation \\Agent};
        \node [below of = Check] 
            (Legend) 
            { \usebox{\legendbox}
            };
        \draw[stream] 
            (Sensor-local-clock) --
            (Fusion);
        \draw[stream] 
            (Sensor-ntp-clock) --
            (Fusion);
        \draw[stream-tuple] 
            (Fusion) --
            (Predict)
            node[midway,right,align=left,xshift=0.5em]
            {Estimated\\parameters\\of Wiener\\process};
        \draw[stream-tuple] 
            (Predict) --
            (Check)
            node[midway,right,align=left,xshift=0.5em]
            {Distribution\\of deviation\\of corrected
            \\local clock\\time from\\the true time};
         \draw[stream-feedback-noarrow,->]
            (Check) -|
            (Sensor-ntp-clock)
            node[midway,right,align=left]
            {Request\\reading}; 
        \draw[stream-feedback-noarrow,->]
            (Check) -|
            (Sensor-local-clock)
            node[midway,left,align=left]
            {Request\\reading};
    \end{tikzpicture}
    \vspace{-2em}
    \caption{Timestamp Sensors Assurance Monitor Agent Network}
    \label{fig:timestamp-network-of-agents}
\end{figure}

Many different algorithms have been used to improve accuracy of timestamps; here we give an example to illustrate the use of the assurance monitor pattern.  Figure~\ref{fig:timestamp-network-of-agents} depicts an agent network for such an application, which uses two sources for time: a local computer clock and an NTP server. The Local clock source agent receives timestamps from the local clock and the NTP clock source agent receives more accurate timestamps from an NTP server. The fusion agent integrates the information it receives from both the local clock and the NTP server to estimate the deviation of the local clock from the true time at every instant.

A prediction agent gets parameters from the fusion agent and predicts
the probability distribution of the deviation as it changes with
time. A check-violation agent determines when the probability of
violating constraints will become unacceptably high and it also keeps
track of the amount of resources that are
available. If sufficient resources (e.g. bandwidth) are available,
then an NTP service is invoked before the constraint is violated. When
resources are not scarce the service is called frequently whereas
when resources are tight, the service is called only when needed.

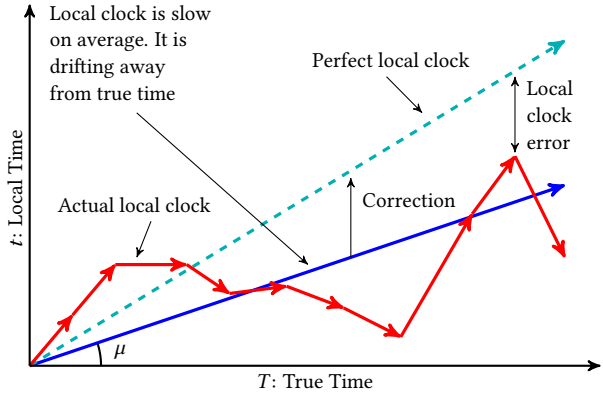
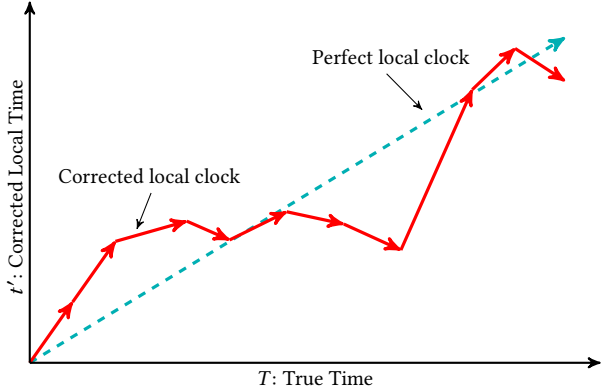
\begin{figure}
    \centering
    \Small
    
    \begin{subfigure}[b]{\linewidth}
        \centering
        \begin{tikzpicture}[scale=0.95]
            \draw [black,thick,domain=0:19] plot ({cos(\x)}, {sin(\x)}) node[right, xshift=0.4em, yshift=-0.45em] {$\mu$};
            \draw[->, very thick, blue] (0,0) -- (7.5,2.5);
            \draw[->, very thick, dashed, TealBlue] (0,0) -- (7.5,4.5);
            \draw[->, very thick,red] (0.0,0.0) -- (0.6,0.7);
            \draw[->, very thick,red] (0.6,0.7) -- (1.2,1.4);
            \draw[->, very thick,red] (1.2,1.4) -- (2.2,1.4);
            \draw[->, very thick,red] (2.2,1.4) -- (2.8,1.0);
            \draw[->, very thick,red] (2.8,1.0) -- (3.6,1.1);
            \draw[->, very thick,red] (3.6,1.1) -- (4.4,0.8);
            \draw[->, very thick,red] (4.4,0.8) -- (5.2,0.4);
            \draw[->, very thick,red] (5.2,0.4) -- (6.2,2.1);
            \draw[->, very thick,red] (6.2,2.1) -- (6.8,2.9);
            \draw[->, very thick,red] (6.8,2.9) -- (7.5,1.5);
            \draw[->] (5.0,4.0) -- (5.5,3.5) node[very near start,yshift=1em] {Perfect local clock};
            \draw[->] (1.7,2.0) -- (1.5,1.5) node[very near start,xshift=-2mm,yshift=1em] {Actual local clock};
            \draw[->] (4.5,1.5) -- (4.5,2.6) node[right,midway,align=left,xshift=0.2em,yshift=0.8em] {Correction};
            \draw[<->] (6.8,2.95) -- (6.8,4.0) node[right,midway,align=left,xshift=0.2em] {Local\\clock\\error};
            \draw[->] (1.5,3.5) -- (3.9,1.4) node[right,near start,align=left,yshift=4.6em,xshift=-6.5em] 
                {Local clock is slow\\on average. It is\\drifting away\\from true time};
            \draw[->, thick, black] (0,0) -- (0,5) 
                node[left,midway,rotate=90,xshift=3em,yshift=0.7em] 
                {$t$: Local Time};
            \draw[->, thick, black] (0,0) -- (8,0) 
                node[below,midway] 
                {$T$: True Time};
        \end{tikzpicture}
        \caption{Local Time Versus True (NTP) Time}
        \label{fig:local-time-versus-true-time}
    \end{subfigure}
    
    \begin{subfigure}[b]{\linewidth}
        \centering
        \begin{tikzpicture}[scale=0.95]
            \draw[->, very thick, dashed, TealBlue] (0,0) -- (7.5,4.5);
            \draw[->, very thick,red] ($(0.0,0.0)    $) -- ($(0.6,0.7*1.2)$);
            \draw[->, very thick,red] ($(0.6,0.7*1.2)$) -- ($(1.2,1.4*1.2)$);
            \draw[->, very thick,red] ($(1.2,1.4*1.2)$) -- ($(2.2,1.4*1.4)$);
            \draw[->, very thick,red] ($(2.2,1.4*1.4)$) -- ($(2.8,1.0*1.7)$);
            \draw[->, very thick,red] ($(2.8,1.0*1.7)$) -- ($(3.6,1.1*1.9)$);
            \draw[->, very thick,red] ($(3.6,1.1*1.9)$) -- ($(4.4,0.8*2.4)$);
            \draw[->, very thick,red] ($(4.4,0.8*2.4)$) -- ($(5.2,0.4*3.9)$);
            \draw[->, very thick,red] ($(5.2,0.4*3.9)$) -- ($(6.2,2.1*1.8)$);
            \draw[->, very thick,red] ($(6.2,2.1*1.8)$) -- ($(6.8,2.9*1.5)$);
            \draw[->, very thick,red] ($(6.8,2.9*1.5)$) -- ($(7.5,1.5*2.6)$);
            \draw[->] (5.0,4.0) -- (5.5,3.5) node[very near start,yshift=1em] {Perfect local clock};
            \draw[->] (1.7,2.35) -- (1.5,1.85) node[very near start,yshift=1em] {Corrected local clock};
            \draw[->, thick, black] (0,0) -- (0,5) 
                node[left,midway,rotate=90,xshift=5.2em,yshift=0.7em] 
                {$t'$: Corrected Local Time};
            \draw[->, thick, black] (0,0) -- (8,0) 
                node[below,midway] 
                {$T$: True Time};
        \end{tikzpicture}
        \caption{Corrected Local Time Versus True Time}
        \label{fig:corrected-local-time-versus-true-time}
    \end{subfigure}
    \caption{Error and Correction of Local Time Versus True (NTP) Time}
    \label{fig:timestamp-times}
\end{figure}

Next, we discuss one of may possible algorithms for the agents in the
application.  The deviation of the true time from the local clock
reading, illustrated in Figure~\ref{fig:timestamp-times}, 
is a random variable $W_{t}$ at time $t$.  This random
variable can be modeled as a Wiener process with drift $\mu$ and
infinitesimal variance $\sigma^{2}$. Then $W_{t+T} - W_{t}$ is a
random variable which has a normal distribution with mean $\mu T$ and
variance $\sigma^{2} T$.
For example, if $\mu = - 0.01$ and $\sigma^{2} = 0.02$ then
after 600 seconds have elapsed, the elapsed time according to
the local clock is a random variable with a normal distribution with
mean $594$ and variance $12$.

The fusion agent estimates $\mu$ and $\sigma$ from a window of a fixed
number of previous measurements. The calculation is based on the
assumptions that the Network Time Protocol (NTP) server is perfectly accurate.  Suppose the
NTP server has been invoked when the local clock reads
$[t_{0}, t_{1}, \ldots , t_{M}]$. Let the NTP server reading be
$T_{j}$ when the local clock is at $t_{j}$. We focus on estimating the
elapsed time between two events rather than on the absolute
time. We use the set of data points
$(t_{j}, T_{j})$ to estimate $\mu$ and $\sigma$ of the Wiener
process. This set of points is a moving window, and now we focus on
the calculation for one window.

The elapsed time local clock times
from the initial point (i.e. the point at which the local clock read $t_{0}$) are: $[(t_{1} - t_{0}), (t_{2} - t_{0}),
\ldots , (t_{M}- t_{0})]$, and the elapsed times as given by the NTP
server at these points are $[(T_{1} - T_{0}), (T_{2} - T_{0}),
\ldots , (T_{M}- T_{0})]$. Plot the points $(t_{j} - t_{0},
T_{j}- T_{0})$ on an $x-y$ plane, and 
obtain a regression line that passes through
the origin. The slope of the regression line is an estimate of
parameter $\mu$ of the Wiener process. Then,
\[
T_{j} - T_{j-1} = \mu (t_{j} - t_{j-1}) + \epsilon_{j}
\]
where $\epsilon_{j}$ is an error term.

Consider a time interval where the local clock reads $t$ at the beginning of
the interval and $t'$ at the end of the interval, where $t_{0} \leq t \leq t'
\leq t_{M}$.
Define the \emph{corrected} elapsed local
time for this interval as:
\[
c(t' - t) = \mu (t' - t)
\]
For the interval in which the local clock reads $t_{j-1}$ at the start
of the interval and $t_{j}$ at the end, we have:
\[
T_{j} - T_{j-1} = c(t_{j} - t_{j-1}) + \epsilon_{j}
\]
and therefore:
\[
 \epsilon_{j}  = (T_{j} - T_{j-1})  - c(t_{j} - t_{j-1}) 
\]
So $\epsilon_{j}$ is the error between the true and corrected local
clock elapsed times for the $j$-th interval.

From the assumptions of a Wiener process, the random
variables of non-overlapping intervals are independent. 
Each $\epsilon_{j}$ is a single sample value of a random variable which has a normal distribution with
zero mean and variance $(T_{j} - T_{j-1})\sigma^{2}$. All these
random variables are independent of each other because the intervals
are non-overlapping.
Hence, the probability density of getting all the sample values $\epsilon_{1},
\epsilon_{2}, \ldots , \epsilon_{M},$ is the product of these probabilty
densities.
We estimate
the variance to be the value $\sigma^{2}$ that maximizes the likelihood of the
set of observations $\epsilon_{j}$, all $j$. The maximum likelihood
estimator for $\sigma^{2}$ is:
\[
\sigma^{2} = \sum_{j} \frac{\epsilon_{j}^{2}}{T_{j} - T_{j-1}}
\]
The estimate of $\sigma$ is passed from the fusion agent to the
prediction agent. 

The prediction agent has the responsibility of
predicting the distribution of the deviation $\epsilon$ of the
corrected local time from the true time, for every instant in the future.  The
check-violation agent has the responsibility of computing the probability that the deviation will
exceed a specified limit for every point in the future. The
check-violation agent then determines the future time after which
this probability exceeds a specified threshold. The NTP server must be
read before this time to satisfy the assurance requirements.

In this example, the prediction agent and check-violation agent are
simple and can be combined into a single, simple agent. Let $t_{M}$ be
the local clock time when the NTP server was read most recently and
let $T_{M}$ be the NTP value. Treat the point $t_{M}, T_{M}$ as the
new origin and compute elapsed time from that point. When
the elapsed time according to the local clock is $t - t_{M}$ then the corrected elapsed time
is $\mu(t - t_{M})$ where $\mu$ is the estimate for the drift. 

The deviation of the corrected time from the true time at a true time
$T$ is a random variable with zero mean and a variance of
$(T-T_{0})\sigma^{2}$ where $\sigma$ is estimated as described
earlier. The agent has access to the local clok but doesn't know the
true time $T$ until it reads an NTP server; so, we use the following
approximation. The deviation of the corrected time from the true time
when the \textit{local clock} is at $t$ is a random variable with zero
mean and a variance of $(t-t_{0})\sigma^{2}$. Computing the local
clock time at which the constraint is violated is now straightforward.

\section{Related and Future Work}

A simple related pattern that has already been identified by the event-based systems community is Event Monitor~\cite{monday2003exploring}. It describes a component that polls a system for state changes and notifies components that are registered with the event monitor, a variant of the observer pattern where the event monitor acts as an intermediary for multiple components interested in monitoring each-other's state. In contrast, assurance monitors predict the occurrence of future events and \emph{act preemptively to prevent unwanted events from occurring}. 
This trait is shared by the Event-based Learning pattern~\cite{paschke2012tutorial}, though assurance monitors combine predictive models with control, that is, a means of enacting remedial actions, as well as means to stage these reactions to account for different event severity levels.

In our future work, we plan to investigate instances of the pattern 
that use more sophisticated prediction agents that rely on extensive 
offline training, such as deep neural networks. These models have 
impressive predictive power but whose accuracy can be hard to reason 
about. 
To address this, we also plan to explore the issue of quantifying the accuracy of predictions used by assurance monitors, which is currently an area of active research~\cite{gehr2018ai2,singh2019abstract}.

\section{Conclusion}

We think a more formal
use of design patterns for event processing will be helpful as more
event processing applications are developed. We described one design
pattern -- the assurance monitor pattern -- for a class of event
processing applications. We described one application, drone
navigation, in detail, and showed how the assurance monitor pattern
can be used in other applications through a second example: time stamping. The pattern,
documentation, and related code are freely available.  Work by the EBS community on design patterns for event
processing, including the development of an open library of patterns,
will be of great service to all who want to develop event-processing
applications.

Why talk about patterns when code that implements the pattern is
available? The code is only one instantiation of the pattern. As we
saw in the timestamp application, some components of the pattern may
become so trivial that they can be combined with other components. In
some applications we may want to execute collections of components in
parallel; for example we may want sensors to execute concurrently and
push their readings to fusion agents. In other applications, we may
want the code to execute within a single process and where the
application pulls data from sensors; we ask the sensors to measure
data only when we need it, as in the case of the GPS sensor for drone
navigation. The range of implementation options is huge -- e.g., single
process versus multiprocess, push versus pull -- but the idea of a
common pattern is helpful in developing each of the implementations.

Another reason for abstracting the pattern is that it helps in
teaching. Indeed, the books on event processing
describe patterns, though sometimes more implicitly than
explicitly. We believe that more formal use of software design
patterns for event processing will help in teaching 
event-based systems and in propagating the research done by the
community.

\bibliographystyle{plain}
\bibliography{main-arxiv}

\end{document}